# A tossed coin as quantum mechanical object

*Alexander M. SOIGUINE*

**Abstract:** Comprehensive and physically consistent model of a tossed coin is presented in terms of geometric algebra. The model clearly shows that there is nothing elementary particle specific in the ½-spin quantum mechanical formalism. It also demonstrates what really is behind this formalism, feasibly reveals the probabilistic meaning of wave function and shows that arithmetic of "packed" objects $|\psi\rangle$ and $\hat{\sigma}_i$ reduces the amount of available information.

> *He who undertakes to deal with questions of natural sciences without the help of geometry is attempting the infeasible.*
> *Galileo Galilei*

## 1. Introduction

Never conclusively ended, Bohr and Einstein's debate on the nature of weirdness of quantum mechanics continued for decades. An interesting hint of possible resolution appeared in 1964 when John Bell [1] formulated mathematical inequality to describe the maximum amount of correlation between properties of two quantum objects if the conditions of realism and localism held. Realism means that any measurable property of an object exists at all times, and its value doesn't depend on someone observing it. In what follows below, the realism will be my main concern.

Let's take a tossed coin and two-value observable, name it "CoinSide", representing the head-tail result of falling coin on a perfectly non-bouncing horizontal plane. Is "CoinSide" a measurable property of the object, tossed coin? If yes, does this measurable property exist at all times not depending on hitting the plane? Isn't it only a result of the measurement which destroys original coin state, falling while rotating? One can get a perception that not everything is simple and straightforward in the realism concept, even in a pure classical experiment.

I will try to formalize and clarify the situation using the tossed coin experiment. The main mathematical working frame will be the even subalgebra $G_3^+$ of elements:

$$so(\alpha, \beta, S) = \alpha + \beta I_S$$

of geometric algebra $G_3$ over Euclidian space $E_3$. $G_3$ has the basis

$$\{1, e_1, e_2, e_3, e_1 e_2, e_2 e_3, e_3 e_1, e_1 e_2 e_3\},$$

where $\{e_i\}$ are the basis (orthonormal) vectors in $E_3$, $\{e_i e_j\}$ - oriented, mutually orthogonal unit value areas spanned by $e_i$ and $e_j$ as edges, $e_1 e_2 e_3$ - unit value oriented volume spanned by ordered edges $e_1$, $e_2$, $e_3$. The basis vector multiplications presenting in these $G_3$ basis elements are geometric products [2], [3]. Subalgebra $G_3^+$ is spanned by $\{1, e_1 e_2, e_2 e_3, e_3 e_1\}$. Variables $\alpha$ and $\beta$ in $so(\alpha, \beta, S)$ are (real[1]) scalars, $I_S$ is a unit size oriented area (lefthanded[2] or righthanded) in an arbitrary given plane $S \subset E_3$.

I explained in detail [4], [5] that elements $so(\alpha, \beta, S) = \alpha + \beta I_S$ only differ from what is traditionally called "complex numbers" by the fact that $S \subset E_3$ is an arbitrary, variable plane and is not the whole space of game. Putting it simply, $\alpha + \beta I_S$ are "complex numbers" depending on $E_2$ embedded into $E_3$. $E_2$ is the space where $S$ belongs. Traditional "imaginary unit" $i$ is just $I_S$ when it is not necessary to specify the plane – everything is going on in one fixed plane, not in 3D world. Fully formal way of using $i$ as a "number" with additional algebraic property $i^2 = -1$ is a source of deeply wrong interpretations in many physical theories.

When $I_S$ is expanded in basis, $I_S = b_1 e_2 e_3 + b_2 e_3 e_1 + b_3 e_1 e_2$, we get the following form of a $G_3^+$ element:

$$\alpha + \beta I_S = \alpha + \beta(b_1 e_2 e_3 + b_2 e_3 e_1 + b_3 e_1 e_2) = \alpha + \beta_1 e_2 e_3 + \beta_2 e_3 e_1 + \beta_3 e_1 e_2, \beta_i = \beta b_i, i = 1,2,3,$$

which is similar to what is traditionally called *quaternion*, with one-to-one correspondence with the three Hamilton's "imaginary" units $i$, $j$, $k$:

$$i \leftrightarrow e_2 e_3, \quad j \leftrightarrow e_1 e_3 \text{ (not } e_3 e_1\text{)}, \quad k \leftrightarrow e_1 e_2, \text{ see [2]}$$

Elements of incomplete form with $\alpha = 0$ are, in terms of geometric algebra, *bivectors,* oriented areas belonging to some $E_2 \subset E_3$ ("pure quaternions").

*Remark 1.1:*

> One can notice that bivectors cannot fully define coin physical state. The state is bivector *plus* instant angle of its rotation. So, coin state should be element of $G_3^+$. Such states have symmetry properties. For example, a coin can also be rotated in its plane around its center axis and such rotation cannot change the result of experiment where we define which side of coin is seen by a fixed observer. The set of all unit value elements of $G_3^+$ (they may also be thought of as elements of unit sphere $S^3$) not changing a unit value bivector $C$ is comprised of all $S^3$ elements with $I_S$ parallel to $C$. Indeed, taking $\alpha + \beta I_S = \alpha + \beta(b_1 e_2 e_3 + b_2 e_3 e_1 + b_3 e_1 e_2)$ and bivector $C = C_1 e_2 e_3 + C_2 e_3 e_1 + C_3 e_1 e_2$, straightforward calculations in geometric algebra terms give the result of rotation (for arithmetic of rotations see [2], [3]) of $C$ by $\alpha + \beta I_S$, $\alpha^2 + \beta^2 = 1$, $b_1^2 + b_2^2 + b_3^2 = 1$:

$$(\alpha - \beta I_S)C(\alpha + \beta I_S) = (\alpha^2 - \beta^2)C - 2\alpha\beta(I_S \times C) - 2\beta^2 I_S(I_S \cdot C), \qquad (1.1)$$

where $I_S \times C = \frac{1}{2}(I_S C - C I_S)$, $I_S \cdot C = \frac{1}{2}(I_S C + C I_S)$. That obviously means that if we rotate bivector with an element of $S^3$ and both have parallel planes, that's $I_S = C$, bivector $C$ remains the same. Nothing changes in our experiment if the coin is also rotating around its center normal!

---

[1] Scalars should always be real. "Complex" scalars are not scalars.
[2] Lefthandedness or righthandedness of an area is the ability to recognize that the area lies on the left (right) while moving along its boundary (counterclockwise or clockwise movement). It actually is a much deeper issue. I am not at the moment considering all of the details.

*Remark 1.2:*

It is necessary to clearly realize that in the geometric algebra sums, like $so(\alpha, \beta, S) = \alpha + \beta I_S$, the addition operation bears the sense "put two things in one bag", not "pour some more wine in a glass"[3].

## 2. A Coin Rotating in 3D

Let's assume very simple type of rotation: the coin is rotating with a constant angular velocity around an axis (the case when axis belongs to the coin plane is shown in Fig.2.1):

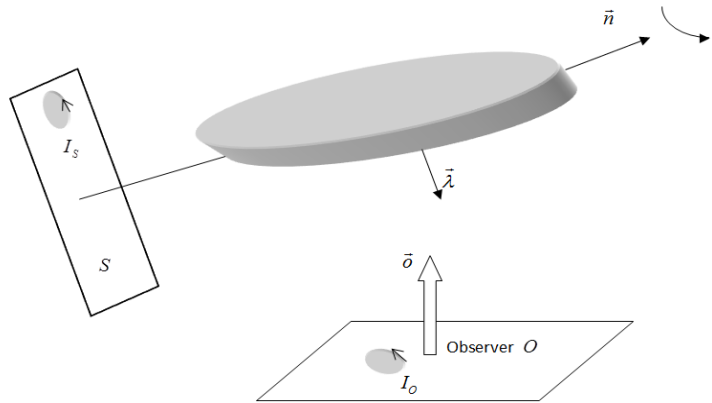

Fig.2.1

---
[3] John W. Arthur [9] used exactly the same explanation of addition of geometrically dissimilar objects as I did in [4] fifteen years earlier.

The unit area (lefthanded, counter clockwise, or righthanded, clockwise) oriented element $I_S$ lies in plane $S$ orthogonal to the axis of rotation $\vec{n}$. An observer looks at the rotating coin as shown in Fig.2.1. The coin sides are, for example, painted in two different colors ("heads" and "tails"). The simplest result of observation at any instance of time is what color the observer does (or does not) see. It is similar to the common way of tossing coin gamble when the coin falls on a horizontal plane (no bouncing assumed, see measurements of B-type below.)

Suppose initial observation of the coin state is $I_C(0)$, a unit bivector. If the coin is rotating with angular velocity $\omega$ around axis $\vec{n}$ then at instant of time $t$ the observable bivector of the coin is [3]:

$$I_C(t) = \exp\left(-\frac{\omega t}{2} I_S\right) I_C(0) \exp\left(\frac{\omega t}{2} I_S\right) \qquad (2.1)$$

Both exponents are full elements of $G_3^+$:

$$\exp\left(\pm \frac{\omega t}{2} I_S\right) = \cos\frac{\omega t}{2} \pm I_S \sin\frac{\omega t}{2}$$

and of unit value:

$$\exp\left(\frac{\omega t}{2} I_S\right)\exp\left(-\frac{\omega t}{2} I_S\right) = \left(\cos\frac{\omega t}{2} + I_S \sin\frac{\omega t}{2}\right)\left(\cos\frac{\omega t}{2} - I_S \sin\frac{\omega t}{2}\right) = 1$$

State as element of $S^3$, or equivalently $G_3^+$, is more thorough thing than quantum mechanical "state", though formally they are very close. At least, I can map one into another. We have:

$$I_C(t) \xleftrightarrow{I_C(0)} \exp\left(\frac{\omega t}{2} I_S\right) \qquad (2.2)$$

State $so_+(\omega, S, t) \equiv \exp\left(\frac{\omega t}{2} I_S\right) = \left(\cos\frac{\omega t}{2} + I_S \sin\frac{\omega t}{2}\right) = \cos\frac{\omega t}{2} + \sin\frac{\omega t}{2}(b_1 e_2 e_3 + b_2 e_3 e_1 + b_3 e_1 e_2)$

satisfies the equation:

$$-I_S \frac{d}{dt} so_+(\omega, S, t) = \frac{\omega}{2} so_+(\omega, S, t) \qquad (2.3)$$

Schrödinger equation for tossing coin!

Relations (2.1) and (2.2) can be viewed in three different, not equivalent ways:

- (2.1) gives transformation $I_C(0) \xrightarrow{\exp\left(\frac{\omega t}{2} I_S\right)} I_C(t)$.
- One way part of (2.2)

$$I_C(t) \xleftarrow{I_C(0)} \exp\left(\frac{\omega t}{2} I_S\right) \qquad (2.2')$$

gives a map of state $\exp\left(\frac{\omega t}{2} I_S\right)$ to observable bivector $I_C(t)$, given the initial not destructed observation $I_C(0)$.

- The other way part of (2.2)

$$I_C(t) \xrightarrow{I_C(0)} \left\{\exp\left(\frac{\omega t}{2} I_S\right)\right\} \qquad (2.2'')$$

gives set of states, the fiber, or level set, of (2.2'), transforming $I_C(0)$ into $I_C(t)$, with arbitrary given plane $S$ the axis of rotation is orthogonal to.

## 3. Hopf Fibrations and Clifford Translations

I will strictly follow the paradigm that (quantum mechanical) evolution equations should be evolution equations for states with explicitly defined "complex" plane.

Transformation (2.1) can be thought of as a (generalized) Hopf fibration, $S^3 \to S^2 : so(\omega, S, t) \to I_C(t)$, generated by bivector $I_C(0)$. Traditional Hopf fibration, see, for example [6], is generated by $I_C(0) = e_2 e_3$, bivector corresponding in our 3D stage to formal "imaginary unit" $i$.

As also follows from the *Remark 1.1*, in usual Hopf fibration, when $I_C(0) = e_2 e_3$, we have for any $G_3^+$ state angle $\varphi$:

$$\exp\left(-\frac{\omega t}{2} I_S\right) e_2 e_3 \exp\left(\frac{\omega t}{2} I_S\right) = \exp\left(-\frac{\omega t}{2} I_S\right) \exp\left(-\frac{\varphi}{2} e_2 e_3\right) e_2 e_3 \exp\left(\frac{\varphi}{2} e_2 e_3\right) \exp\left(\frac{\omega t}{2} I_S\right) \quad (3.1)$$

So, the Hopf fiber is $\exp\left(\frac{\varphi}{2} e_2 e_3\right)$ and we have inside (3.1) Clifford translation

$$\exp\left(\frac{\omega t}{2} I_S\right) \to \exp\left(\frac{\omega t}{2} I_S\right) e^{\frac{\varphi}{2} e_2 e_3} .$$

It is also seen that the state angles in (2.1), (2.2) are two times smaller than object rotation angles. A tossed coin is fermion! We can say: $\frac{1}{2}$-spin fermions are objects in 3D with axial symmetry. Their rotation around the axis of symmetry does not change an observable physical state. In the same way, we can think about 1-spin bosons as objects with spherical symmetry. Rotation of them is one side $G_3^+$ state multiplication when the operand is rotated by the same angle as in state element.

*Remark 3.1:*

The Clifford translations in traditional terms, $z \to ze^{i\varphi}$, are usually considered as acting on unit elements $z \in C^2$ - two dimensional "complex" space, thought of as equivalent to sphere $S^3$.

Common approach of treating $S^3$, which sits in $R^4$, uses stereographic projection and starts with identification of $R^4$ with $C^2$, two dimensional space of "complex" numbers based on their equal dimensionalities:

$$R^4 \supset S^3 = \{z = (z_1, z_2) \in C_2; |z|^2 = \bar{z}_1 z_1 + \bar{z}_2 z_2 = 1\} \quad (3.2)$$

We are considering elements of $S^3$ as states $\alpha + \beta I_S, \alpha^2 + \beta^2 = 1$, and should strictly follow the requirement that if "complex" numbers become involved, their plane must be explicitly defined. When formally using (3.2), a tacit common assumption is that $z_1$ and $z_2$ have the same "complex" plane. Let's make all that unambiguously clear.

Let a state is written as an element of $G_3^+$: $so(\alpha, \vec{\beta}, S) = \alpha + \beta_1 e_2 e_3 + \beta_2 e_3 e_1 + \beta_3 e_1 e_2$

We want to rewrite it as a couple of "complex" numbers, used in (3.2), with an explicitly defined plane. I will initially make assumption that "complex" number plane is one of the spanned by $\{e_2, e_3\}$, $\{e_3, e_1\}$ or $\{e_1, e_2\}$.

For $\{e_2, e_3\}$ we get:

$$so(\alpha, \vec{\beta}) = (\alpha + \beta_1 e_2 e_3) + \beta_2 e_2 e_3 e_1 e_2 + \beta_3 e_1 e_2 = (\alpha + \beta_1 e_2 e_3) + (\beta_3 + \beta_2 e_2 e_3) e_1 e_2 \equiv$$
$$z_1^{2,3} + z_2^{2,3} e_1 e_2$$

For $\{e_3, e_1\}$:

$$so(\alpha, \vec{\beta}) = (\alpha + \beta_2 e_3 e_1) + \beta_3 e_3 e_1 e_2 e_3 + \beta_1 e_2 e_3 = (\alpha + \beta_2 e_3 e_1) + (\beta_1 + \beta_3 e_3 e_1) e_2 e_3 \equiv$$
$$z_1^{3,1} + z_2^{3,1} e_2 e_3$$

And for $\{e_1, e_2\}$:

$$so(\alpha, \vec{\beta}) = (\alpha + \beta_3 e_1 e_2) + \beta_1 e_1 e_2 e_3 e_1 + \beta_2 e_3 e_1 = (\alpha + \beta_3 e_1 e_2) + (\beta_2 + \beta_1 e_1 e_2) e_3 e_1 \equiv$$
$$z_1^{1,2} + z_2^{1,2} e_3 e_1$$

One can notice that all the second members can be written in two ways. For example:

$$z_2^{2,3} e_1 e_2 = (\beta_3 + \beta_2 e_2 e_3) e_1 e_2 = (\beta_2 - \beta_3 e_2 e_3) e_3 e_1.$$

It can be verified that geometrically both give the same bivector.

The bottom line again: if we are speaking about identification $S^3$ and $C^2$, it is necessary to explicitly define which of the three basis (in 3D) "complex" planes we are working in. This plane can also be different from any of basis planes $\{e_i e_j\}$. Instead of basis of three bivectors $\{e_2 e_3, e_3 e_1, e_1 e_2\}$ we can take three unit mutually orthogonal bivectors $\{B_1, B_2, B_3\}$ satisfying the same multiplication rules: $B_1 B_2 = -B_3, B_1 B_3 = B_2, B_2 B_3 = -B_1$. If $S^3 \ni so(\alpha, \beta, S) \equiv \alpha + \beta I_S$ is expanded in basis $\{B_1, B_2, B_3\}$:

$$\alpha + \beta I_S = \alpha + \beta(b_1 B_1 + b_2 B_2 + b_3 B_3) = \alpha + \beta_1 B_1 + \beta_1 B_1 + \beta_1 B_1, \beta_i = \beta b_i,$$

then, for example, taking $B_1$ as "complex" plane we get:

$$\alpha + \beta_1 B_1 + \beta_1 B_1 + \beta_1 B_1 = \alpha + \beta_1 B_1 + \beta_2 B_1 B_3 + \beta_3 B_3 = \alpha + \beta_1 B_1 + (\beta_3 + \beta_2 B_1) B_3 \to$$
$$((\alpha + \beta_1 B_1), (\beta_3 + \beta_2 B_1)) \equiv (z_1, z_2)$$

## 4. Probabilities of Measured Values

As was said in section 2, the two sides of a coin are painted in two different colors. Particularly, the following measurement can be done:

Only the color is defined by a sensor directed as the big arrow in Fig.2.1. This type of a measurement corresponds to the observable used by Bell in his illusive proof of quantum nonlocality [1]:

$$B_{\vec{o}}(\vec{\lambda}) = sign(\vec{\lambda} \cdot \vec{o}),$$

where $\vec{\lambda} = I_C(t) I_3$ and $\vec{o} = I_O I_3$ are dual vectors for the coin instant physical state and the observer bivectors. This type of measurement will be called B-type, honoring J. Bell's efforts to prove that any local realistic extension of quantum mechanics violates experiments.

Let's consider measurements of type B. The problem is to calculate probabilities of two possible results $B_{\vec{o}}(\vec{\lambda}) = sign(\vec{\lambda} \cdot \vec{o})$, where $\vec{\lambda} = I_C(t) I_3$ and $\vec{o} = I_O I_3$ are vectors normal correspondingly to the instant coin plane and fixed observer plane.

Recall the assumptions about physical reality of the tossed coin experiment. The dynamic system under consideration is physical object rotating in 3D. The object is solid disk of negligible thickness. Its orientation at initial instant of time may be unknown. It rotates around unknown axis which is so far supposed to be fixed. The result of measurement is two value observable – which side of the coin the observer observes, formally $sign(\vec{\lambda} \cdot \vec{o})$. No external unknown impacts disturb the rotation. Randomness of the observation result follows from the fact that initial bivector and/or $G_3^+$ state, transforming coin observable bivector, are unknown, are *unspecified variables*[4].

**That's the central point: randomness of observed values appears due to the fact that every observed (measured) value (in the discrete case or a subset in continuous case) corresponds to some particular partition element in space of unspecified variables. Each partition element is fiber (level set)[5] of each of the observable values under function $B_{\vec{o}}(\vec{\lambda}) = sign(\vec{\lambda} \cdot \vec{o})$. Observed value probabilities are (relative) measures of the value fibers under the function of measurement. $G_3^+$ state, "wave function", belongs to (part of) space of unspecified variables.**

The partition of space of unspecified variables as defined above is, generally, different from what is called "sample space" $\Omega$ in the probability space triple $(\Omega, F, P)$ of standard probability theory axiomatics. The sample space there is defined as set of all possible outcomes, the results of a single execution, or a measurement. In the considered experiment, tossing coin, the unspecified variable space is $S^3 \oplus B$. The first component is 3-spere, the set of all $G_3^+$ states. The second component is the set of all (unit) bivectors in 3D – initial coin orientations. That differs from "sample space" of results of execution - two-value observable $sign(\vec{\lambda} \cdot \vec{o})$.

To find the probabilities of the observable values is to find corresponding measures on the product space of states and coin initial bivectors, given each observable value. That means we need to define measures of subsets in $S^3 \oplus B$ that through $(2.2')$ give the two possible results.

Since $I_S = b_1 e_2 e_3 + b_2 e_3 e_1 + b_3 e_1 e_2$, $I_C(0) = C_1 e_2 e_3 + C_2 e_3 e_1 + C_3 e_1 e_2$, we write as before:

$$I_C(t) = (\alpha - \beta I_S)(C_1 e_2 e_3 + C_2 e_3 e_1 + C_3 e_1 e_2)(\alpha + \beta I_S) =$$
$$(\alpha - \beta_1 e_2 e_3 - \beta_2 e_3 e_1 - \beta_3 e_1 e_2)(C_1 e_2 e_3 + C_2 e_3 e_1 + C_3 e_1 e_2)(\alpha + \beta_1 e_2 e_3 + \beta_2 e_3 e_1 + \beta_3 e_1 e_2),$$

where $\beta_i \equiv \beta b_i$. Without losing generality we only will consider measure of the set of $G_3^+$ states which give $I_C(t)$ with normal looking in the hemisphere of basis vector $e_1$.

Since we want to see similarities, parallels with the commonly accepted variant of quantum mechanics, let's explore a brief sideway in the direction of Pauli spinor formalism.

A (pure) state there, associated with a double valued observable, is portrayed with

$$|\psi\rangle = (c_1, c_2)^T,$$

---
[4] I do not want to use highly compromised and ambiguous term "hidden variables".
[5] Recall that fiber of a point $y$ in $Y$ under a function $f : X \rightarrow Y$ is the inverse image of $\{y\}$ under $f$: $f^{-1}(\{y\}) = \{x \in X : f(x) = y\}$

where the components $c_1, c_2$ of the column are "complex" numbers. Our $G_3^+$ states are elements of $G_3^+$ of the form $so(\alpha, \beta, S) \equiv so(\alpha, \vec{\beta}) = \alpha + \beta_1 e_2 e_3 + \beta_2 e_3 e_1 + \beta_3 e_1 e_2, \alpha^2 + \beta^2 = 1$. It was shown above that there exist at least three one-to-one correspondences between elements of $G_3^+$ and couples of "complex" numbers. As was said above, one should strictly follow the requirement that if "complex" numbers are involved their plane must be explicitly defined. Since the Pauli's (and Dirac's) formalism actually makes tacit assumption that "imaginary" $i$ is $e_2 e_3$, we are using

$$so(\alpha, \vec{\beta}) = (\alpha + \beta_1 e_2 e_3) + \beta_2 e_2 e_3 e_1 e_2 + \beta_3 e_1 e_2 = (\alpha + \beta_1 e_2 e_3) + (\beta_3 + \beta_2 e_2 e_3) e_1 e_2 \equiv z_1^{2,3} + z_2^{2,3} e_1 e_2,$$

so:

$$|\psi\rangle = (\alpha + \beta_1 e_2 e_3, \beta_3 + \beta_2 e_2 e_3)^T.$$

As an example, let's get back to (2.3): $-I_S \dfrac{d}{dt} so_+(\omega, S, t) = \dfrac{\omega}{2} so_+(\omega, S, t)$. The state $|\psi\rangle$ corresponding to $so_+(\omega, S, t)$ is:

$$|\psi\rangle = \begin{pmatrix} \cos\dfrac{\omega t}{2} + \beta_1 \sin\dfrac{\omega t}{2} e_2 e_3 \\ \beta_3 \sin\dfrac{\omega t}{2} + \beta_2 \sin\dfrac{\omega t}{2} e_2 e_3 \end{pmatrix}$$

Then equation (2.3), Schrödinger equation for rotating coin, takes the form:

$$-(\beta_1 e_2 e_3 + \beta_2 e_3 e_1 + \beta_3 e_1 e_2)\dfrac{d}{dt}|\psi\rangle = \dfrac{\omega}{2}|\psi\rangle \quad [6]$$

We see again importance of explicit defining the "complex number" plane.

It was shown in [5] that for full formal compatibility of $G_3^+$ formalism, when bivector basis is taken as $\{e_2 e_3, e_3 e_1, e_1 e_2\}$, and Pauli matrix representation it was necessary to reorder and modify a little the Pauli matrices, namely take them as:

$$\hat{\sigma}_1 = \begin{pmatrix} 1 & 0 \\ 0 & -1 \end{pmatrix}, \hat{\sigma}_2 = \begin{pmatrix} 0 & 1 \\ 1 & 0 \end{pmatrix}, \hat{\sigma}_3 = \begin{pmatrix} 0 & i \\ -i & 0 \end{pmatrix}, \quad (i \equiv e_2 e_3)$$

Then the products:

$$\langle\psi|\hat{\sigma}_1|\psi\rangle = \alpha^2 + \beta_1^2 - \beta_2^2 - \beta_3^2, \quad \langle\psi|\hat{\sigma}_2|\psi\rangle = 2(\alpha\beta_3 + \beta_1\beta_2), \quad \langle\psi|\hat{\sigma}_3|\psi\rangle = 2(\beta_1\beta_3 - \alpha\beta_2)$$

---

[6] Basis bivectors should be substituted with Pauli matrices

are exactly the components of usual Hopf fibration received by rotation of $e_2e_3$ through (1.1):

$$(\alpha - \beta I_S)e_2e_3(\alpha + \beta I_S) = (\alpha - \beta(b_1e_2e_3 + b_2e_3e_1 + b_3e_1e_2))e_2e_3(\alpha + \beta(b_1e_2e_3 + b_2e_3e_1 + b_3e_1e_2)) =$$
$$(\alpha - \beta_1e_2e_3 - \beta_2e_3e_1 - \beta_3e_1e_2)e_2e_3(\alpha + \beta_1e_2e_3 + \beta_2e_3e_1 + \beta_3e_1e_2) = (\alpha^2 + \beta_1^2 - \beta_2^2 - \beta_3^2)e_2e_3 +$$
$$2(\alpha\beta_3 + \beta_1\beta_2)e_3e_1 + 2(\beta_1\beta_3 - \alpha\beta_2)e_1e_2, \text{ where } \beta_i \equiv \beta b_i.$$

That also demonstrates that, though we have explicit formulas for $|\psi\rangle \leftrightarrow so(\alpha, \vec{\beta})$, similar arithmetic of "packed" objects $|\psi\rangle$ and $\hat{\sigma}_i$ reduces the amount of available information. Not surprising that it was possible to formally prove [7] that "hidden variables" do not exist in that formalism.

Staying again with usual quantum mechanics two-value spin assumption that $i$ is $e_2e_3$ and with earlier note that it's enough to consider the results looking in hemisphere of the basis vector $e_1$ let's consider the sign of the first component of the rotation result, the $e_1$ looking component of $I_C(t)$, for special case when initial coin bivector is $e_2e_3$.

We know that any extra rotation in the $I_C(0)$ plane does not change the result of the tossing experiment (see *Remark 1.1.*) So, to make calculations looking exactly as in quantum mechanics, let's execute extra rotation defined by $\alpha - \beta_1e_2e_3$:

$$so(\alpha, \vec{\beta}) \to \frac{1}{\sqrt{\alpha^2 + \beta_1^2}}(\alpha - \beta_1e_2e_3)so(\alpha, \vec{\beta})\ ^7.$$ That will eliminate "imaginary" ($e_2e_3$) component from $z_1^{2,3}$:

$$\frac{1}{\sqrt{\alpha^2 + \beta_1^2}}(\alpha - \beta_1e_2e_3)(\alpha + \beta_1e_2e_3 + \beta_2e_3e_1 + \beta_3e_1e_2) = \frac{1}{\sqrt{\alpha^2 + \beta_1^2}}\left[(\alpha^2 + \beta_1^2) + ((\alpha\beta_3 + \beta_1\beta_2) + (\alpha\beta_2 - \beta_1\beta_3)e_2e_3)e_1e_2\right]$$

This corresponds to conventional "wave function":

$$|\psi\rangle = \left(\sqrt{\alpha^2 + \beta_1^2}, \frac{1}{\sqrt{\alpha^2 + \beta_1^2}}((\alpha\beta_3 + \beta_1\beta_2) + (\alpha\beta_2 - \beta_1\beta_3)e_2e_3)\right)^T =$$

$$\left(\sqrt{\alpha^2 + \beta_1^2}, \sqrt{\beta_2^2 + \beta_3^2}\left[\frac{\alpha\beta_3 + \beta_1\beta_2}{\sqrt{\alpha^2 + \beta_1^2}\sqrt{\beta_2^2 + \beta_3^2}} + \frac{\alpha\beta_2 - \beta_1\beta_3}{\sqrt{\alpha^2 + \beta_1^2}\sqrt{\beta_2^2 + \beta_3^2}}e_2e_3\right]\right)^T$$

---

[7] The transformation is equivalent to $e^{-i\theta}|\psi\rangle$ in quantum mechanical terms for appropriate $\theta$. Inverse square root – to stay on $S^3$.

Since $\alpha^2 + \beta_1^2 \leq 1$ and the sum of squares of scalar and bivector components in the last expression is 1 we can formally write:

$$\frac{1}{\sqrt{\alpha^2 + \beta_1^2}}(\alpha - \beta_1 e_2 e_3) so(\alpha, \vec{\beta}) = \sqrt{\alpha^2 + \beta_1^2} + \sqrt{\beta_2^2 + \beta_3^2} I_{S'} = \cos\frac{\theta}{2} + \sin\frac{\theta}{2} I_{S'},\ [8]$$

where $I_{S'}$ does not have $e_2 e_3$ component:

$$I_{S'} = \frac{\alpha\beta_2 - \beta_1\beta_3}{\sqrt{\alpha^2 + \beta_1^2}\sqrt{\beta_2^2 + \beta_3^2}} e_3 e_1 + \frac{\alpha\beta_3 + \beta_1\beta_2}{\sqrt{\alpha^2 + \beta_1^2}\sqrt{\beta_2^2 + \beta_3^2}} e_1 e_2$$

It is orthogonal to $e_2 e_3$ and satisfies: $I_{S'}(e_2 e_3) = -(e_2 e_3) I_{S'} + 2(e_2 e_3) \cdot I_{S'} = -(e_2 e_3) I_{S'}$.

Now we get:

$$\left(\cos\frac{\theta}{2} - \sin\frac{\theta}{2} I_{S'}\right) e_2 e_3 \left(\cos\frac{\theta}{2} + \sin\frac{\theta}{2} I_{S'}\right) = \left(\cos^2\frac{\theta}{2} - \sin^2\frac{\theta}{2}\right) e_2 e_3 - 2\sin\frac{\theta}{2}\cos\frac{\theta}{2} I_{S'} e_2 e_3 =$$

$$\left[(\alpha^2 + \beta_1^2) - (\beta_2^2 + \beta_3^2)\right] e_2 e_3 + 2\sqrt{\alpha^2 + \beta_1^2}\sqrt{\beta_2^2 + \beta_3^2}\left[\frac{\alpha\beta_3 + \beta_1\beta_2}{\sqrt{\alpha^2 + \beta_1^2}\sqrt{\beta_2^2 + \beta_3^2}} e_3 e_1 - \frac{\alpha\beta_2 - \beta_1\beta_3}{\sqrt{\alpha^2 + \beta_1^2}\sqrt{\beta_2^2 + \beta_3^2}} e_1 e_2\right]$$

and the condition of getting positive value of $sign(\vec{\lambda} \cdot \vec{o})$ is:

$$\alpha^2 + \beta_1^2 > \beta_2^2 + \beta_3^2\ [9]$$

or:

$$\alpha^2 + \beta_1^2 > \frac{1}{2} \qquad (4.1)$$

All statrafunctions $so(\alpha, \beta, S) = \alpha + \beta_1 e_2 e_3 + \beta_2 e_3 e_1 + \beta_3 e_1 e_2$ satisfying this condition rotate $e_2 e_3$ in such way that one particular side of coin looks into hemisphere defined by vector $e_1$.

---

[8] Plane of $I_{S'}$ is different from those of $I_S$ or $I_C(0)$. That's, particularly, where standard quantum mechanics is losing information formally writing "imaginary" $i$.

[9] In standard interpretation of QM squared modulus of "complex" components of wave function $|\psi\rangle = (\alpha + \beta_1 i, \beta_3 + \beta_2 i)^T$ are equal to probabilities of finding system correspondingly in states $|1\rangle$ or $|0\rangle$. In our much more detailed theory the condition $\alpha^2 + \beta_1^2 > \beta_2^2 + \beta_3^2$ defines relative measure of all $G_3^+$ states transforming initial bivector $e_2 e_3$ into one with positive component in $e_2 e_3$ plane.

It was shown in the *Remark 3.1* how to construct three mappings of a $G_3^+$ state to $C^2$, corresponding to selection of one particular "complex" plane $e_i e_j$. Right now, to calculate measures of level sets on $S^3$, it will be convenient to use the following parameterization. If

$$so(\alpha, \beta, S) \to ((\alpha + \beta_{k \neq i,j} e_i e_j), (\beta_j + \beta_i e_i e_j)) \equiv (z_1, z_2), \quad |z_1|^2 + |z_2|^2 = 1$$

then Hopf coordinates $(\eta, \xi_1, \xi_2)$, $0 \leq \eta \leq \frac{\pi}{2}$, $0 \leq \xi_i \leq 2\pi$ can be used to get:

$$z_1 = (\cos \xi_1 + e_i e_j \sin \xi_1) \sin \eta$$

$$z_2 = (\cos \xi_2 + e_i e_j \sin \xi_2) \cos \eta$$

Currently, $e_i e_j = e_2 e_3$ and (4.1) inequality becomes:

$$|z_1| = |\sin \eta| > \frac{\sqrt{2}}{2},$$

that means $\frac{\pi}{4} < \eta < \frac{\pi}{2}$. Corresponding integration with measure element on $S^3$ in Hopf coordinates, $\sin \eta \cos \eta \, d\eta \, d\xi_1 \, d\xi_2$, gives the area:

$$\int_{\frac{\pi}{4}}^{\frac{\pi}{2}} \sin \eta \cos \eta \, d\eta \int_0^{2\pi} d\xi_1 \int_0^{2\pi} d\xi_2 = \pi^2,$$

half of the $S^3$ full surface value $2\pi^2$. **So, exactly half of all states returns positive value to observable $sign(\vec{\lambda} \cdot \vec{o})$, in the case when initial coin plane is parallel to $e_2 e_3$.**

**The result is wonderful. It explains in absolutely clear way why we have fifty–fifty head/tail tossed coin observation if axis of coin rotation is randomly and uniformly distributed in 3D.**

Easy to show that calculations with (4.1) changed to opposite $\alpha^2 + \beta_1^2 < \frac{1}{2}$ give the same result: coin sides are identical.

Let's calculate relative measures of states returning positive values to $sign(\vec{\lambda} \cdot \vec{o})$ when initial state is $e_3 e_1$. Eliminating the $e_3 e_1$ component from $z_1^{3,1}$, as was done for the $e_2 e_3$ case just to make everything similar to common quantum mechanics, is not necessary. The observation of $e_3 e_1$ in a $G_3^+$ coin state is:

$$(\alpha - \beta_1 e_2 e_3 - \beta_2 e_3 e_1 - \beta_3 e_1 e_2) e_3 e_1 (\alpha + \beta_1 e_2 e_3 + \beta_2 e_3 e_1 + \beta_3 e_1 e_2) =$$
$$2(-\alpha\beta_3 + \beta_1\beta_2) e_2 e_3 + (\alpha^2 + \beta_2^2 - \beta_1^2 - \beta_3^2) e_3 e_1 + 2(\alpha\beta_1 + \beta_2\beta_3) e_1 e_2$$

Positive value of the considered observable means $(-\alpha\beta_3 + \beta_1\beta_2) > 0$. This is the condition on the set of states returning necessary result when initial coin orientation is $e_3 e_1$, so we take:

$$\alpha + \beta_1 e_2 e_3 + \beta_2 e_3 e_1 + \beta_3 e_1 e_2 = (\alpha + \beta_2 e_3 e_1) + \beta_3 e_3 e_1 e_2 e_3 + \beta_1 e_2 e_3 = (\alpha + \beta_2 e_3 e_1) + (\beta_1 + \beta_3 e_3 e_1) e_2 e_3$$

Condition $(-\alpha\beta_3 + \beta_1\beta_2) > 0$ in terms of $z_1 = \alpha + \beta_2 e_3 e_1$, $z_2 = \beta_1 + \beta_3 e_3 e_1$, $\alpha = \cos\xi_1 \sin\eta$, $\beta_1 = \cos\xi_2 \cos\eta$, $\beta_2 = \sin\xi_1 \sin\eta$, $\beta_3 = \sin\xi_2 \cos\eta$ can be written in coordinates $(\eta, \xi_1, \xi_2)$ as:

$$\cos\xi_1 \sin\eta \sin\xi_2 \cos\eta < \cos\xi_2 \cos\eta \sin\xi_1 \sin\eta,$$

$$\sin(\xi_1 - \xi_2) > 0,$$

$$0 < \xi_1 - \xi_2 < \pi,$$

and we get integral for the area on the $S^3$ surface giving states with the observation result $(-\alpha\beta_3 + \beta_1\beta_2) > 0$:

$$\int_0^{\frac{\pi}{2}} \sin\eta \cos\eta \, d\eta \int_0^{2\pi} d\xi_2 \int_{\xi_2}^{\pi+\xi_2} d\xi_1 = \pi^2$$

Very interesting! Probability of a particular value of observable $sign(\vec{\lambda} \cdot \vec{o})$ does not depend is initial coin orientation $e_2 e_3$ or $e_3 e_1$ (obviously, the same for $e_1 e_2$). This fact will in future be used for the Stern-Gerlach experiment explanations.

## 5. Conclusions

It was shown that:

- Elements of $G_3^+$ are "complex numbers" when the latter are more thoroughly considered as existing in three dimensions.
- Quantum mechanical "wave function" should be considered as an element of $G_3^+$, not two dimensional complex valued state $|\psi\rangle$.
- Actually, a mapping exists between quantum mechanical objects in a pure state $|\psi\rangle = (c_1, c_2)^T$ and classical tossed coin $G_3^+$ states, though arithmetic of "packed" objects $|\psi\rangle$ and $\hat{\sigma}_i$ reduces the amount of available information.

- Probabilities of the two-value experiment are naturally calculable from fiber measures in the space of unspecified variables – $G_3^+$ states returning bivector coin state (3D space orientation) as measured observable.

Further works will apply the approach to magnetic dipoles and Stern-Gerlach experiment.